\documentclass[preprint,prm,showkeys,superscriptaddress]{revtex4-2}

\usepackage{graphicx}
\usepackage{amsmath,amsfonts,amssymb}
\usepackage{nicefrac}
\usepackage[colorlinks=true,linkcolor=black,citecolor=blue,urlcolor=blue]{hyperref}

\begin{document}

  \title{Dislocations and plasticity of KTaO$_3$ perovskite\\modeled with a new interatomic potential}

  \author{Pierre~Hirel\footnote{Corresponding author: \url{pierre.hirel@univ-lille.fr}}}
  \affiliation{Univ. Lille, CNRS, INRAE, Centrale Lille, UMR 8207 - UMET - Unit\'{e} Mat\'{e}riaux et Transformations, F-59000 Lille, France}
  
  \author{Franck Junior Kakdeu Yewou}
  \affiliation{Univ. Lille, CNRS, INRAE, Centrale Lille, UMR 8207 - UMET - Unit\'{e} Mat\'{e}riaux et Transformations, F-59000 Lille, France}
  \affiliation{Univ. Gustave Eiffel, Univ. Paris Est Créteil, CNRS, MSME, F-77454 Marne-la-Vallée, France}
  
  \author{Jiawen Zhang}
  \affiliation{Department of Mechanical and Energy Engineering, Southern University of Science and Technology, Shenzhen, China}
  
  \author{Wenjun Lu}
  \affiliation{Department of Mechanical and Energy Engineering, Southern University of Science and Technology, Shenzhen, China}
  
  \author{Xufei Fang}
  \affiliation{Institute for Applied Materials, Karlsruhe Institute of Technology, Kaiserstr. 12, 76131, Karlsruhe Germany}
  
  \author{Philippe Carrez}
  \affiliation{Univ. Lille, CNRS, INRAE, Centrale Lille, UMR 8207 - UMET - Unit\'{e} Mat\'{e}riaux et Transformations, F-59000 Lille, France}
  
  \date{\today}

\begin{abstract}
   Potassium tantalate KTaO$_3$ is a cubic, paraelectric perovskite ceramic that exhibits surprising ductility at room temperature as most recently reported. Much like strontium titanate (SrTiO$_3$), plastic deformation is accommodated by dislocations gliding in $\{110\}$ planes. In this work we propose a new interatomic potential for KTaO$_3$, and apply it to model dislocations with $\langle 110\rangle$ Burgers vector. We demonstrate that dislocations dissociate, and finely characterize their core structure and Peierls potential. Dislocations of edge character can carry a positive or negative electric charge, but we show that charge-neutral configurations are energetically more favorable. We also perform high-resolution electron microscopy to validate our simulation methodology.  Comparing our results with other ductile perovskites, we confirm KTaO$_3$ to be ductile, but stiffer than SrTiO$_3$.

\end{abstract}

\maketitle

\section*{Introduction}

In the last decade, tantalate perovskites ATaO$_3$ where A is an alkali metal (A=Li,Na,K,Rb,Cs,Fr), have attracted a lot of attention owing to their broad technological applications, such as energy storage \cite{Zhao2017,Yin2018,Edalati2020}, photoluminescence and photo-catalysis \cite{Hu2009,Sumedha2021,Sumedha2022}, spintronics \cite{Hussain2021,Gupta2022,Hussain2024}, or wastewater treatment \cite{Patra2023}. While in semiconductors it is widely considered that doping elements are vital and dislocations are fatal to functional properties \cite{Queisser1998}, in oxide ceramics defects engineering is seriously considered as a way to enhance functional properties \cite{Nakamura2003,Gao2018,Armstrong2021,Fang2024}. For instance, KTaO$_3$ nano-flowers offer an increased surface over volume ratio and enhance photo-catalysis performance \cite{Sumedha2021,Sumedha2022}.

However, studies of dislocations and their effects on functional properties remain scarce. In strontium titanate SrTiO$_3$, plastic deformation was linked to the activation of $\{110\}$ slip planes \cite{Brunner2001a,Sigle2006}, and dislocations were shown to enable switching from insulating to conductive state, associated with oxygen diffusion along dislocation lines \cite{Szot2006,Rodenbucher2023}. In ferroelectric potassium niobate KNbO$_3$ single crystals, ductile deformation leads to interactions between dislocations and domain walls \cite{Hirel2015}. The third perovskite demonstrated to be ductile is potassium tantalate KTaO$_3$, by recent bulk compression tests, independently by Khayr \cite{Khayr2024} and Fang \cite{Fang2024KTO}. Plastic deformation of bulk ceramics at room temperature remains an intriguing phenomenon, and raises questions about elementary mechanisms underlying it.

Atomic-scale simulations can provide insight into the elementary mechanisms of plasticity. To that end, large atomic systems are required, especially in perovskite materials where dislocations have large dissociation distances \cite{Hirel2012}. Such large systems are out of range of \emph{ab initio} calculations, and less computationally demanding methods must be used. Existing interatomic potentials were successfully applied to dislocation modeling in SrTiO$_3$ \cite{Hirel2012,Marrocchelli2015,Hirel2025}, however no reliable potential exist for KTaO$_3$.

\begin{table}[b]
  \caption{Parameters for the rigid-ion potential (RIP) used in this study (Eq.~\ref{pot}). Ion charges are $q_\text{K}=0.6e$, $q_\text{Ta}=3e$, and $q_\text{O}=-1.2e$. Parameters for K$-$O and O$-$O interactions are from ref.~\cite{Pedone2006}, those for Ta$-$O were fitted in this study. The cut-off radius is 15~\AA .}
  \begin{tabular}{lcccc}
  \hline
         &  $D_{ij}$ &  $a_{ij}$     & $r_0$  & $C_{ij}$ \\
         &    (eV)   & (\AA $^{-2}$) & (\AA ) & (eV$\cdot$\AA $^{12}$) \\
 \hline \hline
K$-$O    &  0.011612 & 2.062605 & 3.305308 & 5.0  \\
Ta$-$O   &  0.008508 & 3.006309 & 2.719528 & 1.0  \\
O$-$O    &  0.042395 & 1.379316 & 3.618701 & 22.0 \\
 \hline
  \end{tabular}
  \label{potparam}
\end{table}

In this work we perform the first atomic-scale simulations of dislocations in cubic KTaO$_3$ perovskite. To that end we fit a new interatomic potential and validate it against \emph{ab initio} data. Then we model dislocations of $\langle 110\rangle$ Burgers vector, associated with plastic deformation in related perovskites \cite{Sigle2006,Hirel2012}. Dislocations of screw character dissociate in their $\{110\}$ glide plane. Dislocations of edge character also dissociate, and are shown to be preferentially charge-neutral. High-resolution electron microscopy observation of deformed KTaO$_3$ samples confirm the dissociation both qualitatively and quantitatively, and validate predictions made with our interatomic potential. Comparison with SrTiO$_3$ and KNbO$_3$ sheds light on elementary mechanisms associated with room-temperature plasticity in perovskite ceramics.


\begin{table*}[!t]
\begin{tabular}{llllcc}
\hline   
         &         Expt.          &               DFT           & DFT  & SMP              & Modified RIP \\
         &                        &                             &      & (Sepliarsky \cite{Sepliarsky2005}) & (this study)   \\
\hline \hline
$a_0$    &  3.95 \cite{Weber2018}& 3.99 (LDA \cite{Xu2017}) & 4.06 (PBE \cite{Shigemi2006})     &   3.99 ($-$)        &  4.00 ($-0.2$\%) \\
         &  3.99 \cite{Hu2022}   & 4.06 (GGA \cite{Xu2017}) & 4.00 (PBESol \cite{Benrekia2012}) & &     \\
         &                        &  3.99 (HSE06 \cite{Xu2017}) & 3.99 (PBESol \cite{Bouafia2013})  &    &  \\
\hline   
$C_{11}$ &  431 \cite{Weber2018}  &  492 (LDA \cite{Xu2017})   & 451 (PBESol \cite{Benrekia2012})  &      429 ($-0.5$\%) &  429 ($-0.5$\%)  \\
         &                        &  420 (GGA \cite{Xu2017})   & 474 (PBESol \cite{Bouafia2013})   &    &     \\
         &                        &  470 (HSE06 \cite{Xu2017}) &           &    &     \\
\hline   
$C_{12}$ &  103 \cite{Weber2018}  &  83 (LDA \cite{Xu2017})   & 81 (PBESol \cite{Benrekia2012})  &      105        ($+1.9$\%) &  104 ($+1$\%)   \\
         &                        &  77 (GGA \cite{Xu2017})   & 63 (PBESol \cite{Bouafia2013})   &    &     \\
         &                        &  87 (HSE06 \cite{Xu2017}) &            &    &     \\
\hline   
$C_{44}$ &  109 \cite{Weber2018}  &  102 (LDA \cite{Xu2017})  & 102 (PBESol \cite{Benrekia2012})  &      105 ($+3.7$\%)        &  104  ($-4.6$\%)  \\
         &                        &  100 (GGA \cite{Xu2017})  & 197 (PBESol \cite{Bouafia2013})   &    &     \\
         &                        &  112 (HSE06 \cite{Xu2017}) &           &    &     \\
\hline   
$A$      & 0.665 \cite{Weber2018} &  0.50 (LDA \cite{Xu2017})  & 0.55 (PBESol \cite{Benrekia2012}) &     0.65 ($-2.3$\%)        &  0.64 ($-3.8$\%) \\
         &                        &  0.58 (GGA \cite{Xu2017})  & 0.96 (PBESol \cite{Bouafia2013})  &    &     \\
         &                        &  0.58 (HSE06 \cite{Xu2017}) &          &    &     \\
\hline
$\gamma _\text{APB}$ &            &  0.28 (present study)      &           & 0.86                  & 0.50 \\
\hline
\end{tabular}
\caption{Ambient-pressure properties of cubic KTaO$_3$ computed with the modified rigid-ion potential (present study), and compared with those computed with the shell-model potential (SMP) by Sepliarsky \emph{et al.} \cite{Sepliarsky2005}, and with experimental and density functional theory (DFT) data from literature. Lattice parameter $a_0$ is given in \AA , elastic constants $C_{ij}$ in GPa, and (110) APB energy $\gamma _\text{APB}$ in J$\cdot$m$^{-2}$. Zener anisotropy factor is computed as $A = 2 C_{44} / (C_{11} - C_{12})$. Experimental values were obtained at ambient temperature. Numbers in parenthesis give the deviation with respect to experimental values.}
\label{bulkprop}
\end{table*}

\section{Methods and models}

\subsection{Interatomic potential}

Potassium tantalate (KTaO$_3$ or KTO) crystallizes in the cubic perovskite structure at all temperatures. Classical molecular statics (MS) calculations are carried out with LAMMPS \cite{Plimpton1995}. Interactions between ions are described by means of a rigid ion potential (RIP) that includes the Coulomb interaction, a short-range Morse function, and a repulsive term:

\begin{equation}
  U(r_{ij}) = \frac{q_i q_j}{r_{ij}} + D_{ij} \{ \left[ 1 - \exp (a_{ij} (r_{ij}-r_0)) \right] ^2 - 1 \} + \frac{C_{ij}}{r_{ij}^{12}}
  \label{pot}
\end{equation}

where $q_i$ and $q_j$ are the electric charges of ions, $r_{ij}$ the distance between them, and $D_{ij}$, $a_{ij}$, $r_0$ and $C_{ij}$ are adjustable parameters. Coulomb interaction is evaluated by means of the particle-particle particle-mesh (pppm) algorithm with a relative accuracy of 10$^{-6}$. Relaxation is performed by means of the conjugate-gradients (CG) algorithm, until total energy is converged within $10^{-9}$~eV/atom.

We use the parameters proposed by Pedone et al. \cite{Pedone2006}, where the electric charge of an ion is an integer multiple of the partial charge $q = 0.6e$ where $e$ is the elementary charge, i.e. $q_\text{K}=0.6e$ and $q_\text{O}=-1.2e$. Parameters for O$-$O interactions were fitted to reproduce the basic properties (lattice parameter, elastic constants) of $\alpha$-quartz (SiO$_2$), while parameters for K$-$O interactions were fitted to potassium silicate K$_2$Si$_2$O$_5$, as explained in the original article \cite{Pedone2006}.

The original potential by Pedone does not contain parameters for Ta$^{5+}$ ions. To maintain consistency with the original potential, and to ensure charge neutrality of the KTaO$_3$ composition, we set the electric charge of Ta ions to $q_\text{Ta} = 5 q_\text{K} = 3e$. The parameter $C_{ij}$ in Eq.~\ref{pot} affects only the very short-ranged term ($C_{ij}/r^{12}$), which takes importance only in high-temperature molecular dynamics according to the authors \cite{Pedone2006}. Hence we choose to equal it to 1, i.e. the same value as for similar ions in the B site of a perovskite structure (e.g. Ti or Si). In the end, only parameters $D_{ij}$, $a_{ij}$, and $r_0$, in the Morse function remain to be fitted. Target observables are the lattice constant $a_0$, and elastic constants $C_{11}$, $C_{12}$ and $C_{44}$, and the fitting is performed using GULP \cite{Gale1997}. Cubic symmetry is enforced during the fitting procedure, and a cut-off of 15~\AA \ is used. 

Potential parameters, including those obtained for Ta$-$O after the fitting procedure, are reported in Table~\ref{potparam}. We then plugged this potential into LAMMPS to relax the structure, and compute the resulting lattice parameter and elastic constants. Results are presented in Table~\ref{bulkprop}, and compared to experimental and DFT data from literature. It can be seen that DFT calculations produce different values depending on the choice of exchange-correlation functional, the HSE06 functional being the closest match to experimental values \cite{Xu2017}. Our rigid-ion potential reproduces experimental values with good accuracy concerning both lattice parameter, elastic constants, and Zener anisotropy, confirming the robustness of the fitting procedure. We also note that bulk physical properties produced by our RIP are very similar to the ones obtained with the shell-model potential (SMP) by Sepliarsky \cite{Sepliarsky2005}.

\begin{figure*}[ht!]
  \centering
  \includegraphics[width=\linewidth]{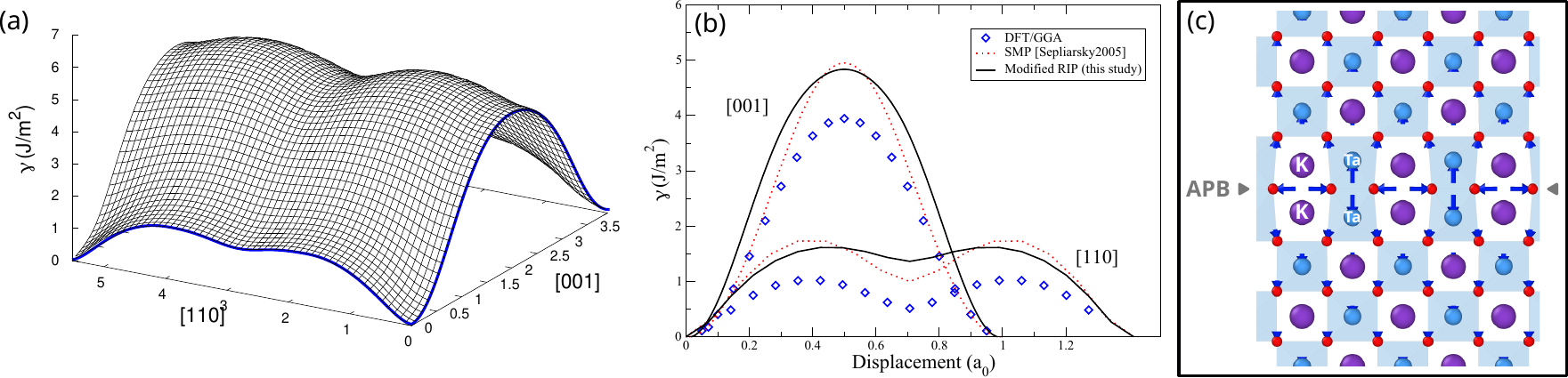}
  \caption{\small{Energy density $\gamma$ (in J$\cdot$m$^{-2}$) of generalized stacking faults (GSF) in $(\bar{1}10)$ plane in KTaO$_3$. (a) Complete $\gamma$ surface computed with the rigid-ion potential (RIP) modified in this study. (b) Values of $\gamma$ along [110] and [001] computed with DFT/GGA (blue diamonds), Sepliarsky shell-model potential (red dotted lines), and modified rigid-ion potential (black curves). (c) Ionic displacements in the vicinity of the APB when ions are allowed to fully relax. Potassium ions are displayed as large violet spheres, tantalum as medium blue spheres, oxygen as small red spheres, and TaO$_6$ octahedra are shown in transparent blue. Blue arrows show atomic displacements magnified by a factor~6 for better visualization.}}
\label{Fig1}
\end{figure*}

\subsection{Dislocation dipole setup}

To model dislocations, a unit cell of cubic KTaO$_3$ is duplicated to form a $80\times 80\times 1$ supercell, of approximate size $440 \times 440 \times 4$~\AA $^3$, counting 63,600~ions. Crystal orientation is $X=[110]$, $Y=[\bar{1}10]$, $Z=[001]$. A dipole of dislocations is introduced with their Burgers vector parallel to $X=[110]$. Screw dislocations are introduced by applying displacements according to elastic dislocation theory \cite{Hirth1982}. Edge dislocations are introduced by removing a half-plane of atoms, as detailed below. Energy is minimized by means of the conjugate-gradients algorithm, with respect to both ion positions and cell geometry. In the following, relative energies are expressed in meV per angström of dislocation line, to facilitate comparison between dislocations with different periodicities.

Atomic systems are constructed with Atomsk \cite{Hirel2015Atomsk}. Visualization is performed with VESTA \cite{Momma2011} and OVITO \cite{Stukowski2010}. To visualize defects, the centro-symmetry criterion \cite{Kelchner1998_PRB} is computed using cations only (K and Ta) which form a bcc-like structure, so its value is naught for ions in perfect environment, and positive for ions in a defect (typically dislocations in the present study). The criterion is not computed for oxygen ions.

\subsection{Computation of Peierls barriers}

To evaluate Peierls energy barriers we use the nudged elastic band (NEB) method \cite{Jonsson1998}. The computation uses 9 to 16 replicas, and a spring force of 1~eV/\AA \ is applied between replicas to keep them equidistant. The calculation is considered converged when the maximum force on a replica falls below 10$^{-2}$~eV/\AA . Then the climbing image method \cite{Henkelman2000} is applied so that a replica reaches the saddle point. The activation energy is then computed as the difference in total energy between the saddle point and the initial replica.

\subsection{Charge density analysis}

To estimate the distribution of electric charges around dislocations, we use an atomic-to-continuum approach detailed in our previous work \cite{Hirel2025}. To summarize, ions (point charges) are replaced by 2-D Gaussian functions in the plane normal to the dislocation line. In the bulk Gaussian functions overlap and compensate, while in defects the overlap is imperfect and a net charge arises. Integrating the charge density in the region surrounding the defect gives an estimation of its electric charge.

\subsection{Experiments and high-resolution electron microscopy}

Single-side polished, undoped KTiO$_3$ (001) single crystals ($5\times 5 \times 1$~mm$^3$) were sourced from Hefei Ruijing Optoelectronics Technology Co., Ltd. (Anhui, China). These crystals were produced using the flame-melting technique, yielding a low initial dislocation density (about 10$^{10}$~m$^{-2}$). To introduce dislocations at room temperature, scratching tests were performed using an Rtec MFT2000 wear testing machine (UK) equipped with a 3~mm-diameter Al$_2$O$_3$ (ruby) spherical indenter, similar to previous experiments \cite{Fang2024KTO}.

For microstructure analysis of the dislocations, transmission electron microscope (TEM) lamella specimens were prepared using a dual-beam focused ion beam (FIB) in a scanning electron microscope (SEM, Helios Nanolab 600i, FEI, Hillsboro, USA). The TEM samples, taken along the scratching direction, were lifted out from the center of the wear track after 5~cycles of scratching, allowing for better visualization of the dislocation structures from different observation perspectives. Scanning TEM (STEM) images were captured using a TEM instrument (double aberration-corrected FEI Titan Themis G2) operating at 300~kV.  In the high-angle annual dark field scanning TEM (HAADF-STEM) images, a probe semi-convergence angle of 17~mrad and inner and outer semi-collection angles ranging from 38 to 200~mrad were employed.

\section{Generalized stacking faults}

Before modeling dislocations, we computed the energy density of generalized stacking faults (GSF) in a $(\bar{1}10)$ plane of KTO. Crystal orientation is $X=[110]$, $Y=[\bar{1}10]$, $Z=[001]$, and the system has a size of $1\times 6\times 1$. The top half of the system along $Y$ is shifted rigidly along $X$ of $Z$ to form a stacking fault. Ions are constrained to relax only along the $Y$ direction normal to the fault. Atomistic simulations are performed with the SMP by Sepliarsky \cite{Sepliarsky2005}, and with the rigid-ion potential (RIP) that we modified above. For comparison, we also carry out high-accuracy first-principles calculations using the VASP code \cite{Kresse1996} which enables Density Functional Theory (DFT) simulations using a plane-wave basis set, combined with the Projector Augmented Wave (PAW) method \cite{Blochl1994}. Exchange-correlation energy is approximated with the Perdew–Burke–Ernzerhof Generalized Gradient Approximation (PBE-GGA) functional \cite{Perdew1996,KresseJoubert1999}. Consistent with the system size, a plane wave cut-off energy of 500~eV, and $k$-points sampling $4\times 1\times 6$ \cite{Monkhorst1976} are used.

Fig.~\ref{Fig1}a shows the complete $\gamma$-surface computed with the modified RIP. The main features are comparable to similar calculations performed in SrTiO$_3$ \cite{Hirel2010,Klomp2023}: the lowest energies are found along [110], with a local minimum at $\nicefrac{1}{2} [110]$ corresponding to the metastable anti-phase boundary (APB). The second lowest energy path is found along [001], but corresponds to unstable states.

Fig.~\ref{Fig1}b compares energies along [110] and [001] computed with different methods. DFT predicts a single maximum $\gamma = 4$~J$\cdot$m$^{-2}$ along [001], and a double hump along [110], with a metastable anti-phase boundary (APB) with a (constrained) energy about $0.71$~J$\cdot$m$^{-2}$. Both interatomic potentials over-estimate energies along [110], the SMP by a factor of 1.4, and our modified Pedone potential by a factor of 2 approximately.

Having confirmed the existence of a metastable APB, we removed the constraints on atoms and allowed them to relax in all directions. The final, fully relaxed, APB energy drops drastically with all three methods: it becomes $\gamma _\text{APB} = 0.28$~J$\cdot$m$^{-2}$ with DFT, 0.86~J$\cdot$m$^{-2}$ with Sepliarsky SMP, and 0.50~J$\cdot$m$^{-2}$ with our modified RIP. This large gain in energy comes with significant relaxation, as illustrated in Fig.~\ref{Fig1}c. This probably occurs because repulsion between oxygen ions is not sufficiently screened by pairs of K$^+$ cations, while pairs of Ta$^{5+}$ cations attract oxygen ions to them. In addition, Ta ions repel each other and shift away from the APB, so that they are not in the same (110) plane as K ions. As a result from these relaxations, TaO$_6$ octahedra are not regular any longer, but are significantly distorted in the APB (Fig.~\ref{Fig1}c). Both interatomic potentials correctly reproduce this relaxation. It is worth noting that despite these distortions, unit cells in the APB remain paraelectric, since Ta ions remain at the center of oxygen octahedra. Also, such complex relaxation was not found in SrTiO$_3$, where instead octahedra remain mostly regular in the APB \cite{Hirel2010,Hirel2012}.

Despite its accuracy in bulk material properties, the SMP by Sepliarsky \emph{et al.} has three main disadvantages. The first is its complexity: the presence of shells doubles the number of particles per unit cell, making the computation of interactions more time consuming. It also complicates the relaxation of defects, the additional degrees of freedom giving rise to local energy minima where relaxation algorithms can be trapped, and sometimes cause numerical instability. The second problem has to do with electric charges of particles. The Sepliarsky SMP uses charges that are not multiple of one another ($q_\text{K} = 0.82e$, $q_\text{Ta} = 4.84e$, $q_\text{O} = -1.89e$), meaning that partial Schottky defects, such as K$_2$O or Ta$_2$O$_5$ clusters, would not be charge-neutral. Worse than that, the Ta core is assigned a negative charge ($-2.9878e$) while its shell, supposed to mimic the electronic cloud, has a large positive charge ($+7.8265e$) \cite{Sepliarsky2005}. Not only are these values non-physical, but they also make it much more difficult to make sense of charged defects. Finally, the third drawback is that after full relaxation, its prediction for the APB energy is further away from the DFT value than our more simple modified RIP.

On the contrary, our RIP uses partial charges that are compatible with charge-neutral vacancy clusters (see Table~\ref{potparam}), and is computationally less demanding than a SMP. We acknowledge that the APB energy of 0.50~J$\cdot$m$^{-2}$ is higher than the DFT value by a factor of 1.78, but it is closer to DFT than any existing interatomic potential. Assuming inverse proportionality between APB energy and dissociation distance as predicted from elastic theory of dislocations \cite{Hirth1982}, $\gamma _\text{APB} = \mu b^2 / (2\pi \kappa d)$ (where $\kappa=1-\nu$ and $1$ for screw and edge dislocations respectively), this factor of 1.78 can be accounted for, similarly to what was done in SrTiO$_3$ \cite{Hirel2012}. Qualitatively, our RIP offers a good description of the $\gamma$-surface as a whole, without any discontinuity. In the following, we use our modified RIP to model dislocation core structures.

\begin{figure*}[ht!]
  \centering
  \includegraphics[width=0.85\linewidth]{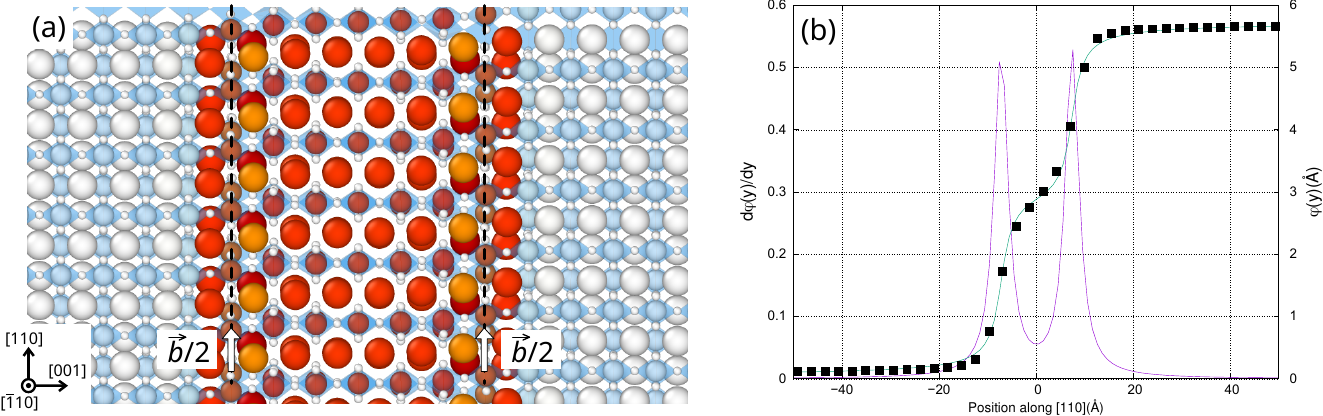}
  \caption{\small{(a) Atomic core structure of the screw dislocation with Burgers vector $\vec{b} = [110]$. After relaxation, it spreads preferentially in the $(\bar{1}10)$ plane, where it dissociates into two collinear partial dislocations separated by an APB. K ions are displayed as large spheres, Ta as medium spheres, oxygen as small spheres, and TaO octahedra are displayed in transparent blue. Ions are colored according to their centro-symmetry criterion (see Methods). (b) Disregistry $\phi$ associated with the screw dislocation (black dots), and its derivative (purple curve). The latter shows two maxima giving the position of the two partial dislocations.}}
\label{Fig2}
\end{figure*}

\section{$\langle 110\rangle \{\bar{1}10\} $ screw dislocations}

\begin{figure}[!t]
  \centering
  \includegraphics[width=\linewidth]{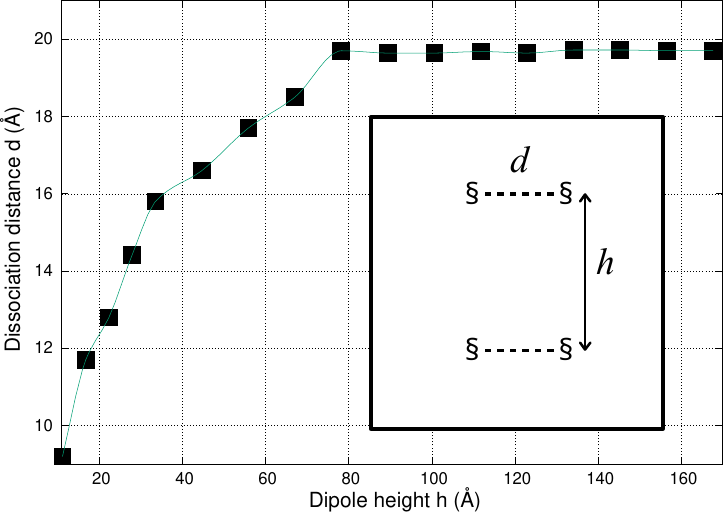}
  \caption{\small{Evolution of dissociation width $d$ as function of the height $h$ of a dipole of [110] screw dislocations. The inset shows a schematic of the dipole configuration.}}
\label{Fig3}
\end{figure}

\subsection{Atomic core structure}

Two screw dislocations are introduced with opposite Burgers vectors $\pm [110]$. After relaxation they spontaneously dissociate in the $(\bar{1}10)$ plane, as expected from the previous $\gamma$-surface calculation. Fig.~\ref{Fig2} shows the resulting atomic structure, revealing that two partials are separated by an APB, however octahedra are severely distorted near the cores of partial dislocations. Computation of the disregistry (black squares  in Fig.~\ref{Fig2}) confirms that the partials have collinear Burgers vectors $\nicefrac{1}{2}[110]$. The derivative $d\phi / dx$ (purple curve) shows two peaks, where each partial has a width of about 5~\AA , and their dissociation width is 19.4~\AA . The fact that the derivative does not vanish between the two peaks shows that partial dislocations overlap, causing a distortion of the APB. Accounting for the over-estimation of the APB energy by a factor of 1.78 (see previous section), our calculations predict that the dissociation width should actually be around $w^\text{§} \approx 34$~\AA .

\subsection{Dependence on dipole height}

In SrTiO$_3$ there is evidence that for screw dislocations in a dipole configuration, the dissociation width depends on the dipole height, i.e. on the distance between two dislocations of opposite Burgers vectors. This effect was first observed experimentally by Castillo-Rodr\'iguez and Sigle \cite{Castillo-Rodriguez2010}, and then confirmed by atomistic simulations \cite{Hirel2012}. The reason is that when dislocations are close to one another, partials with opposite Burgers vectors $\pm \nicefrac{1}{2}[110]$ attract each other very strongly, causing a reduction of the dissociation width. When dislocations become separated enough, this effect becomes negligible with respect to the repulsion of collinear partials and to the stacking fault energy, and dislocations can be considered almost isolated from each other.

To study this effect in KTaO$_3$, we varied the height of the dislocation dipole $h$, and after relaxation, measured the dissociation width $d$ of each dislocation. Results are reported in Fig.~\ref{Fig3}. It can be seen that for $h < 80$~\AA , dissociation width strongly depends on the dipole height. For the smallest height $h=16$~\AA \ (i.e. about 4 unit cells), dissociation width is reduced to 8~\AA . Below this height, the two screw dislocations are so constricted and attract each other so strongly, that they cross-slip in (110) planes and annihilate, leaving only perfect crystal. For heights $h>80$~\AA , dissociation width of each dislocation reaches a plateau and does not depend on the dipole height any longer. This means that dislocations are separated enough and can relax into their optimal atomic structure. These simulations indicate that dislocations must be separated by a height $h\geq 80$~\AA \ to relax independently and to study individual dislocations.

\subsection{Peierls barrier}

\begin{figure}[!t]
  \centering
  \includegraphics[width=\linewidth]{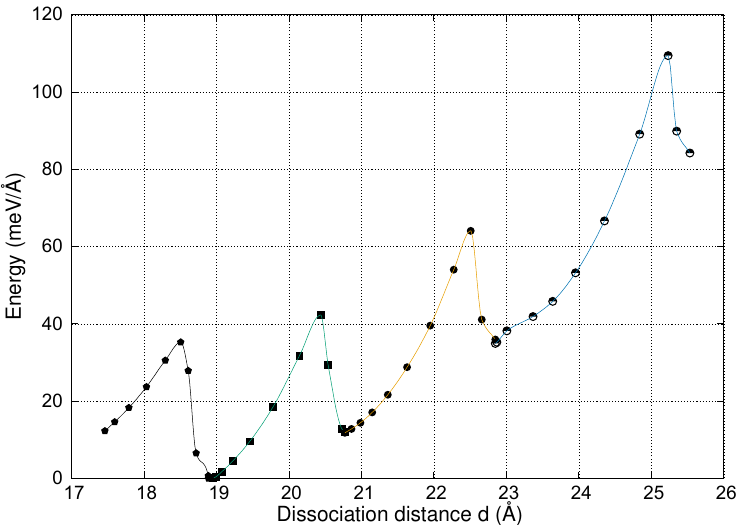}
  \caption{\small{
 Relative energy of a screw [110] dislocation as function of the dissociation distance $d$ between the two partials. Several configurations are found to be metastable. Energies are given with respect to the most stable configuration. }}
\label{Fig4}
\end{figure}

To study the motion of the screw dislocation through the lattice, we construct screw dislocations with different imposed dissociation widths, and relax each of these configurations independently. This produces metastable configurations where the dissociation width ranges from 4 to 8 lattice constants $a_0$. Then we compute the energy barrier between metastable configurations, i.e. the Peierls barrier, by means of the NEB method. The four final energy paths are represented in Fig.~\ref{Fig4}.

Overall, as the dissociation distance $d$ increases the system's total energy increases. This is due to the contribution of the stacking fault energy, which increases with $d$. The most stable configuration is found at $d=19.4$~\AA \ or $5a_0$, consistent with the previous results. Other distances $d=4$, 6, 7 and 8 are only metastable. Between these configurations, the Peierls barrier is roughly the same, and we find an Peierls energy about 45~meV$\cdot$\AA $^{-1}$.

\section{$\langle 110\rangle \{\bar{1}10\}$ edge dislocations}

\subsection{Electrically charged edge dislocation}

\begin{figure*}[ht!]
  \centering
  \includegraphics[width=\linewidth]{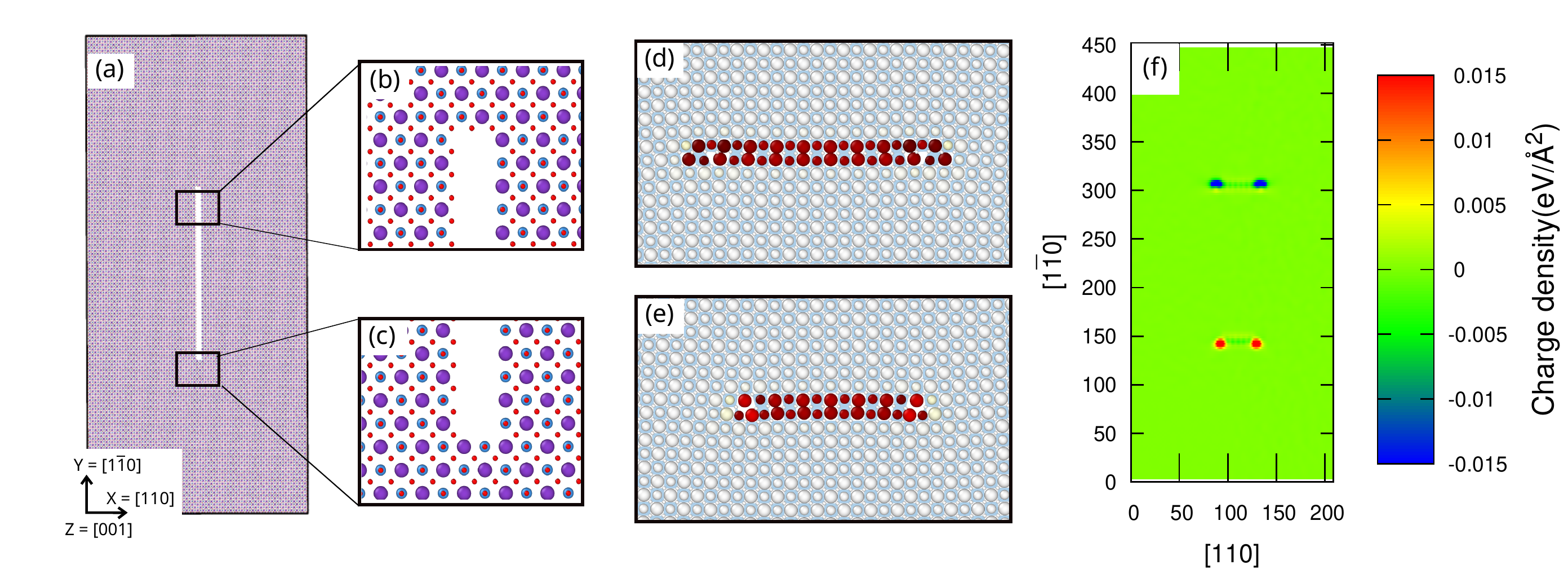}
  \caption{\small{Construction of a dipole of charged [110] edge dislocations (same color code as Fig.~\ref{Fig2}). (a) A ribbon of unit cells is removed, leading to formation of two nonequivalent $(\bar{1}10)$ surfaces, (b) oxygen rich, and (c) oxygen poor. (d) After relaxation the cut closes and two dislocations are formed, both dissociated, but with two different core structures. (e) Charge analysis reveals that the two dislocations carry opposite charges $\pm q$, and that the charge is localized on partial dislocations.}}
\label{Fig5}
\end{figure*}

A dipole of edge dislocations is constructed by removing cells from a perfect crystal, following a procedure already applied to SrTiO$_3$ \cite{Marrocchelli2015,Klomp2023,Hirel2025}. In the middle of a $80\times 80\times 1$ supercell, a ribbon of width [110] is removed, thus leaving a void region or gap in the material. One extremity is terminated with a plane of oxygen ions, therefore a negatively charged $(\bar{1}10)$ surface, and the other by a plane of KTaO$_2$ stoichiometry hence positively charged (Fig.~\ref{Fig5}). The cell is compressed along the $X=[110]$ direction to compensate for the artificially negative pressure induced by the gap, and facilitate further relaxation. First, ions are allowed to relax only along the $X$ direction in order to close the gap that was formed. Then constraints are removed and ions are allowed to relax in all directions. Finally, cell geometry is optimized so that no stress is applied to the structure.

Fig.~\ref{Fig5} shows the resulting configurations. Two [110] edge dislocations form as expected, both dissociated in their $(\bar{1}10)$ glide plane. However their atomic core structures are different, one containing extra oxygen atoms with respect to the other. This is confirmed by charge analysis, one dislocation carrying a charge $+q$ (red areas in Fig.~\ref{Fig5}) and the other a charge $-q$ (blue areas). Computation of disregistries show that dissociation widths are very close, 33.9~\AA \ for the positively charged dislocation, and 34.8~\AA \ for the negatively charged. Accounting for the factor 1.78 as above, we estimate the dissociation width of charged dislocations to be about 61~\AA .

Although electrically charged dislocations can be constructed and stabilized in simulations, they are expected to have a very high energy contribution. The longer the dislocation segment the higher the energy cost. In the end, long charged dislocations are not expected to remain in the material, at least not without a mechanism that would compensate their charge.

\subsection{Charge-neutral edge dislocation}

\begin{figure*}[ht!]
  \centering
  \includegraphics[width=\linewidth]{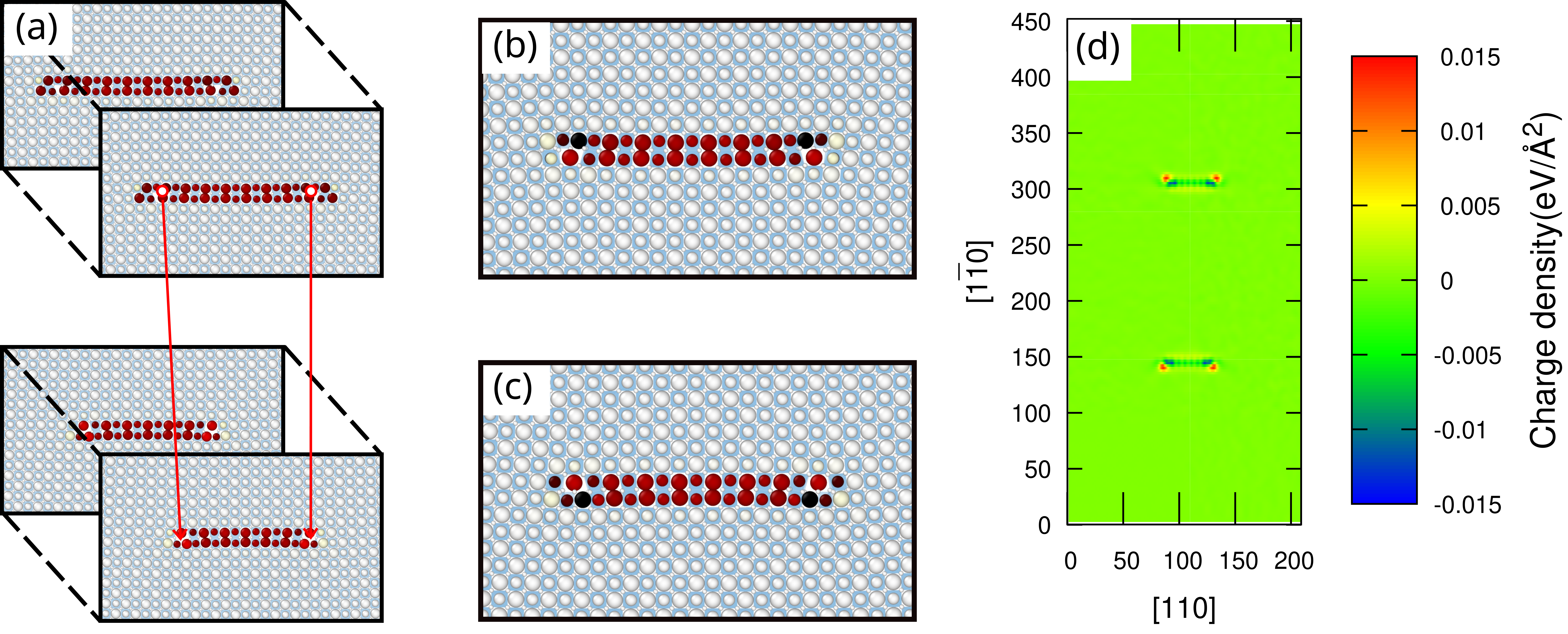}
  \caption{\small{Method for neutralizing the charge of [110] edge dislocations. (a) The system is duplicated along dislocation lines $Z=[001]$, and half oxygen columns are displaced from negatively-charged partials into positively-charged partials. (b,c) After relaxation, two equivalent [110] edge dislocations are obtained. Along each partial dislocation line, one oxygen site out of two is vacant. (d) Charge analysis reveals that each partial dislocation is charge-neutral.}}
\label{Fig6}
\end{figure*}

To obtain charge-neutral dislocations, ions can be moved from one charged dislocation to the other. Oxygen vacancies having the highest mobility, with a migration barrier around 0.9~eV in unstrained KTO \cite{Xi2017}, it is sensible to move oxygen ions from the negatively charged dislocation into the positive one. In our previous work on SrTiO$_3$, we have shown that displacing an entire oxygen column prevents dissociation, and leads to the formation of compact, sessile dislocations \cite{Hirel2025}. Instead, it is more favorable to duplicate the system along dislocation lines, and to move half an oxygen column from each negative partial dislocation into its positive counterpart, as illustrated in Fig.~\ref{Fig6}. After relaxation, one obtains two equivalent edge dislocation cores that are still dissociated, where each partial dislocation contains a half-filled oxygen column (Fig.~\ref{Fig6}b). Charge analysis confirms that each partial is charge-neutral (Fig.~\ref{Fig6}c).

Compared to the system containing charged dislocations, the one with charge-neutral dislocations is lower in energy by 22.7~eV. This confirms that charge-neutral dislocations are much more favorable than their charged counterparts, just like in SrTiO$_3$ \cite{Hirel2025}. The dissociation distance is also smaller, 31.5~\AA , which translates into $w^\perp = 60$~\AA \ after accounting for the factor of 1.78. This is expected because partial dislocations carrying the same charge impose elastic and electrostatic repulsion on one another, while in charge-neutral partials the electrostatic contribution effectively vanishes, thus reducing dissociation width.

\subsection{HRTEM observations}

\begin{figure*}[!t]
  \centering
  \includegraphics[width=\linewidth]{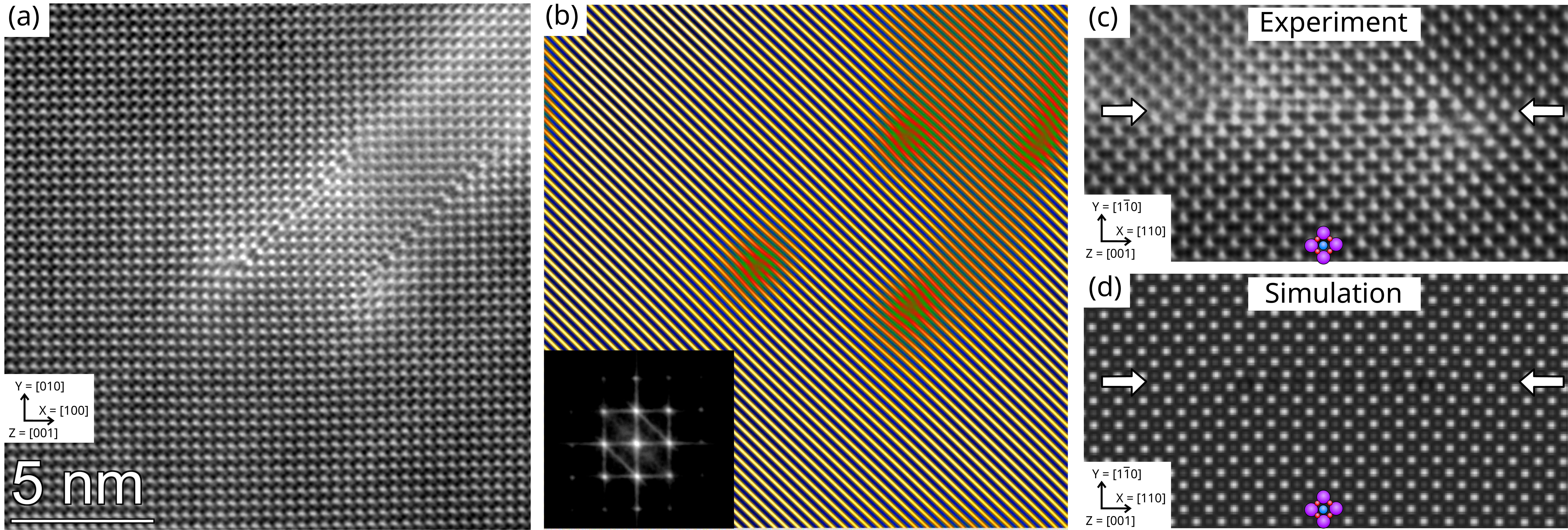}
  \caption{ 
  \small{(a) HAADF-STEM image of a dipole of $\pm [110]$ edge dislocations. (b) IFFT of the region in (a), showing the extra planes. (c) Zoom-in on one of the edge dislocations observed in (a), rotated by 45$^\circ$. The slip plane of the dislocation is indicated with white arrows. The unit cell of KTaO$_3$ is represented for reference (colored spheres). (d) Same edge dislocation as simulated in Fig.~\ref{Fig6}, where atoms were replaced by 2-D Gaussian functions to facilitate comparison with experiments. Dissociation width is smaller than in experiment because the interatomic potential over-estimates the APB energy (see text).} }
\label{Fig7}
\end{figure*}

To further validate our simulation methodology, we employed HAADF-STEM imaging, combined with FFT and inverse FFT (IFFT), to analyze the core structures of edge dislocations after five cycles of scratching. Fig.~\ref{Fig7}a shows a dipole of dislocations of edge character. One may note that the occurrence of dipoles is likely attributed to significant dislocation multiplication occurring after five cycles of scratching. As observed in molecular static simulations, $\langle 110\rangle$$\{\bar{1}10\}$ edge dislocations show a clear tendency to dissociate in $\{\bar{1}10\}$ with a stacking fault corresponding to the APB described previously. IFFT image further highlights the partial dislocations as shown in Fig.~\ref{Fig7}b. According to Fig.~\ref{Fig7}b, we estimate a separation width between partial dislocations of 70~\AA . This leads to a upper limit estimate of the APB extension within the dislocation core (Fig.~\ref{Fig7}c), nevertheless fully consistent with the value predicted by our simulations (60~\AA ).

\subsection{Peierls barrier}

\begin{figure}[!b]
  \centering
  \includegraphics[width=\linewidth]{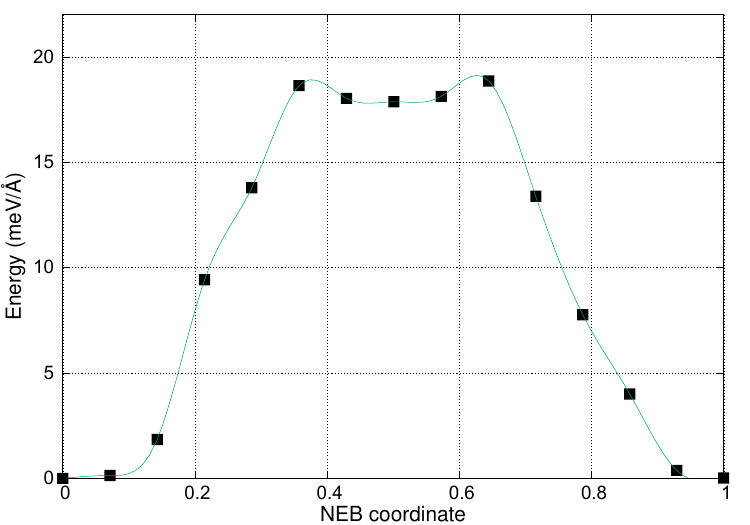}
  \caption{\small{Minimum energy path for the motion of a charge-neutral and dissociated [110] edge dislocation in KTaO$_3$.}}
\label{Fig8}
\end{figure}

To study the motion of the charge-neutral edge dislocation, we apply simple shear to the simulation cell. After each shear increment of 0.1\% the system is fully relaxed, and the procedure is repeated until the edge dislocation (i.e. both partials) has moved by one lattice vector. Then, all strain is removed and the system is fully relaxed again, yielding the final configuration. The NEB method is used to obtain the minimum energy path between the initial and final configurations.

Fig.~\ref{Fig8} shows the resulting Peierls energy barrier. It does not show a single maximum, but instead a double-hump shape, although the local minimum is very shallow. The total energy barrier is about 40~meV$\cdot$\AA$^{-1}$. Looking at the atomic configurations, this energy path corresponds to a motion of the trailing partial which slightly compresses the dislocation into a local energy minimum, rapidly followed by motion of the leading partial that restores the most favorable dissociation width. This motion is quite subtle, but markedly different from SrTiO$_3$, where edge dislocations move as a whole and the Peierls barrier shows a single energy minimum \cite{Hirel2025}. It would be desirable to investigate how this energy path changes with applied stress, and what are the consequences for plastic deformation.

\section{Discussion}

\subsection{Comparison with SrTiO$_3$ and KNbO$_3$}

After strontium titanate (SrTiO$_3$ or STO) and potassium niobate (KNbO$_3$ or KNO), potassium tantalate KTaO$_3$ is the third perovskite that was shown to exhibit ductility at room temperature \cite{Khayr2024,Fang2024KTO}. Although at room temperature bulk STO and KTO are both cubic ($Pm\bar{3}m$ space group) while KNbO$_3$ is orthorhombic and ferroelectric ($Amm2$), in terms of composition and physical properties KTO is more closely related to KNO than to STO (see Table~\ref{comp}). Concerning dislocations, direct comparison is difficult because different interatomic potentials were used in literature (e.g. STO was modeled either with Thomas \cite{Hirel2012} or Pedone \cite{Hirel2025} potentials, and KNO was modeled with Sepliarsky SMP \cite{Hirel2015}), however we attempt to highlight the main trends.

On one hand, a property that the three materials have in common is their polar $\{110\}$ planes, that can be terminated with ABO$_2$ (hence a charge $+2$) or oxygen ions ($-2$). This means that the same electric charges arise when constructing edge dislocations, and the same charge compensation mechanisms can be applied. As a result, charge-neutral edge dislocations have similar structures in STO \cite{Hirel2025}, KNO \cite{Hirel2015} and KTO (present study). This explains why glide in $\{110\}$ is favored in all three materials, and why they all exhibit a certain ductility.
Furthermore, the $\gamma$-surface of KTO (Fig.~\ref{Fig1}) shares all the main features of the one in STO \cite{Hirel2010}, in particular a metastable APB along [110]. As a result, it is not surprising that the same $\langle 110\rangle \{\bar{1}10\}$ slip systems are activated in KTO \cite{Fang2024}, KNO \cite{Hirel2015} and STO \cite{Sigle2006}, and that [110] dislocations dissociate.

On the other hand, the three materials differ in their elastic constants, as shown in Table~\ref{bulkprop}). In particular, the shear modulus $\mu$ is larger in KTO and KNO, which makes these two materials stiffer than STO. This trend is reflected in the Peierls energy barrier that we obtained: for edge dislocations, $V _\text{P} ^\perp \approx 40$~meV$\cdot$\AA $^{-2}$ in KTO, and $5.4$~meV$\cdot$\AA $^{-2}$ in STO \cite{Hirel2025}. This is consistent with recent nano-indentation experiments by Fang and co-workers, who estimated a critical resolved shear stress $\tau _\text{CRSS} \approx 137$~MPa in KTO at room temperature \cite{Fang2024}, larger than that of STO $\tau _\text{CRSS} \approx 90$~MPa in similar experiments \cite{Javaid2018}. Thus KTO appears to be less ductile than STO.


\begin{table}[!t]
  \begin{tabular}{lrrr}
  \hline              &     SrTiO$_3$          &   KNbO$_3$             & KTaO$_3$ \\
  \hline \hline 
  C$_{11}$            & 255 \cite{Hirel2025}   & 426 \cite{Xu2017}      & 429  \\
  C$_{12}$            & 126 \cite{Hirel2025}   &  84 \cite{Xu2017}      & 104  \\
  C$_{44}$            & 126 \cite{Hirel2025}   & 104 \cite{Xu2017}      & 104  \\
  $\mu$               &  90                    & 133                    & 130  \\
  $\gamma_\text{APB}$ & 0.925 \cite{Hirel2010} & 0.50 \cite{Hirel2015}  & 0.28 \\
  $w^\text{§}$        &  39 \cite{Hirel2012}   &  30 \cite{Hirel2015}   &  34  \\
  $w^\perp$           &  28.6 \cite{Hirel2025}  &  60 \cite{Hirel2015}   &  61  \\
  \hline
  \end{tabular}
  \caption{Elastic constants (GPa) for the three perovskite materials discussed here. Shear modulus (GPa) is computed as $\mu = \sqrt{C_{44} \cdot (C_{11}-C_{12})/2}$. APB energies $\gamma _\text{APB}$ are given in J$\cdot$m$^{-2}$, dislocation dissociation widths $w$ in \AA , for dislocations of screw (§) and edge ($\perp$) characters. Values for KTaO$_3$ come from the present work. }
  \label{comp}
\end{table}

\subsection{Ductile to brittle transition}

A remarkable feature of SrTiO$_3$ is that around 1000~K it looses its ductility and becomes brittle \cite{Brunner2001a}, a transition that appears linked to the compaction of edge dislocation segments leading eventually to climb dissociation \cite{Hirel2016b,Hirel2025}. However such a ductile-brittle transition was not observed in KNbO$_3$, which remains ductile up to 1200~K, with a flow stress that decreases above 800~K \cite{Hirel2015}. This is probably due to dissociation distances being larger in KNO ($w^{\perp} \approx 60$~\AA) than in STO (28.6~\AA ), making their compaction much more difficult. If this hypothesis is correct then it would also apply to KTaO$_3$, i.e. the large dissociation width (61~\AA ) would also prevent compaction of edge segments, and KTO would not have a ductile to brittle transition, but would remain ductile even at high temperature. This scenario could be confirmed by high-temperature experiments, and/or by further atomic-scale simulations of a glide-to-climb transformation, as was done in SrTiO$_3$ \cite{Hirel2025}. We are confident that the interatomic potential that we propose here can be used to study such mechanisms.

\subsection{Charge of edge dislocations}

As expected, our simulations show that a system containing charged dislocations has a large excess energy compared to a system containing charge-neutral dislocations. As a result, one does not expect long dislocations carrying a lineic charge to exist in the material.

Just like in SrTiO$_3$ \cite{Hirel2025}, negatively charged (oxygen-rich) dislocations are inconsistent with oxygen vacancies that are expected to be present at any finite temperature. Therefore we argue that the formation of oxygen-rich dislocations would be very unlikely. In the end, the role of negatively charged dislocations in plastic deformation can be dismissed.

This means that [110] edge dislocations are either charge-neutral, or carry a positive electric charge. Due to the large electrostatic energy cost, it is reasonable to assume that long dislocation lines with positive lineic charge are not favorable. Instead, our calculations suggest that charge-neutral dislocations are much more favorable, and can easily glide through the lattice. If such a dislocation encounters an oxygen vacancy, it can locally become positively charged, while still remaining glissile. Later this charge can be balanced by emitting an oxygen vacancy, restoring the charge-neutral core. This indicates that edge dislocations should keep a good mobility even at high temperature, and in a wide range of oxygen partial pressures. The exact mechanisms of dislocation-vacancy interaction would require further simulations to be characterized in detail.

\subsection{$\langle 100 \rangle$ dislocations}

STO, KNO and KTO also differ in ionicity. In STO the A and B sites carry formal charges of $+2$ (Sr) and $+4$ (Ti), while in KNO and KTO the charges are $+1$ (K) and $+5$ (Ta or Nb). Interestingly enough, this has no incidence on the formal charge of $\{110\}$ planes, as explained above, but it does change the charge associated with $\{100\}$ planes.

In STO, $\{100\}$ planes consist of a stacking of charge-neutral SrO and TiO$_2$ planes. Thus, edge dislocations with Burgers vector $\textbf{b}=\langle 100\rangle$ can have the stoichiometry SrO or TiO$_2$, but both variants would be charge-neutral. On the contrary, in KTO and KNO, $\{100\}$ planes terminated with KO carry a charge of $-1$, while those of TaO$_2$ or NbO$_2$ termination carry a charge of $+1$. As a consequence, $\langle 100\rangle$ edge dislocations with such terminations would not be charge neutral, and charge compensation mechanisms would have to be applied. It would be valuable to investigate the stoichiometry and charge of $\langle 100 \rangle$ dislocations, as they can play a role in grain boundaries or at (100)//(100) interfaces. The charge and charge-compensation mechanisms associated with $\langle 100\rangle$ dislocations may also influence electronic and ionic conduction in these materials.

\section*{Conclusion}

We propose a new interatomic potential for potassium tantalate KTaO$_3$, and demonstrate its applicability to the modeling of dislocations belonging to $\{110\}$ slip planes. Similarly to SrTiO$_3$, both edge and screw dislocations in KTaO$_3$ are dissociated and glide relatively easily through the lattice. Edge dislocations are preferentially charge-neutral, but remain glissile even if they become positively charged (oxygen deficient), a sign that the material should remain ductile in a wide range of temperatures and oxygen partial pressures. This work opens new pathways for investigating elementary mechanisms of plastic deformation and diffusion in KTaO$_3$ and related materials.

\section*{Acknowledgments}

DFT and atomistic simulations were carried out at the HPC center at the DGDNum of University of Lille. W. L. acknowledges the funding from the Shenzhen Science and Technology Program (Grant No. JCYJ20230807093416034). X. F. acknowledges the funding from the European Union (ERC Starting Grant, Project MECERDIS, grant No. 101076167).

\section*{Author Contributions}

\textbf{P.H.:} conceptualization, interatomic potential fitting, supervision, methodology, data curation, formal analysis, writing - original draft, review and editing.
\textbf{F.J.K.Y.:} atomistic simulations, data curation, formal analysis, writing - review and editing.
\textbf{J.Z.:} HRSTEM analysis, writing - review and editing.
\textbf{W.J.:} experiments, HRSTEM observations and analysis, supervision, data curation, writing - review and editing.
\textbf{X.F.:} supervision, data curation, writing - review and editing.
\textbf{P.C.:} DFT calculations, supervision, data curation, formal analysis, writing - review and editing.


\begin{thebibliography}{10}
\expandafter\ifx\csname url\endcsname\relax
  \def\url#1{\texttt{#1}}\fi
\expandafter\ifx\csname urlprefix\endcsname\relax\def\urlprefix{URL }\fi
\expandafter\ifx\csname href\endcsname\relax
  \def\href#1#2{#2} \def\path#1{#1}\fi

\bibitem{Zhao2017}
L.~Zhao, Q.~Liu, J.~Gao, S.~Zhang, J.~F. Li, Lead-free antiferroelectric silver
  niobate tantalate with high energy storage performance, Advanced Materials 29
  (8 2017).
\newblock \href {https://doi.org/10.1002/adma.201701824}
  {\path{doi:10.1002/adma.201701824}}.

\bibitem{Yin2018}
J.~Yin, Y.~Zhang, X.~Lv, J.~Wu, Ultrahigh energy-storage potential under low
  electric field in bismuth sodium titanate-based perovskite ferroelectrics,
  Journal of Materials Chemistry A 6 (2018) 9823--9832.
\newblock \href {https://doi.org/10.1039/c8ta00474a}
  {\path{doi:10.1039/c8ta00474a}}.

\bibitem{Edalati2020}
K.~Edalati, K.~Fujiwara, S.~Takechi, Q.~Wang, M.~Arita, M.~Watanabe,
  X.~Sauvage, T.~Ishihara, Z.~Horita, Improved photocatalytic hydrogen
  evolution on tantalate perovskites CsTaO$_3$ and LiTaO$_3$ by strain-induced
  vacancies, ACS Applied Energy Materials 3 (2020) 1710--1718.
\newblock \href {https://doi.org/10.1021/acsaem.9b02197}
  {\path{doi:10.1021/acsaem.9b02197}}.

\bibitem{Hu2009}
C.~C. Hu, C.~C. Tsai, H.~Teng, Structure characterization and tuning of
  perovskite-like NaTaO$_3$ for applications in photoluminescence and
  photocatalysis, Journal of the American Ceramic Society 92 (2009) 460--466.
\newblock \href {https://doi.org/10.1111/j.1551-2916.2008.02869.x}
  {\path{doi:10.1111/j.1551-2916.2008.02869.x}}.

\bibitem{Sumedha2021}
H.~N. Sumedha, M.~Shashank, F.~A. Alharthi, M.~S. Santosh, B.~M. Praveen,
  G.~Nagaraju, Synthesis of novel pseudo-capacitive perovskite nanostructured
  flowerlike KTaO$_3$ for lithium ion storage, International Journal of Hydrogen
  Energy 46 (2021) 28214--28220.
\newblock \href {https://doi.org/10.1016/j.ijhydene.2021.06.045}
  {\path{doi:10.1016/j.ijhydene.2021.06.045}}.

\bibitem{Sumedha2022}
H.~N. Sumedha, M.~Shashank, S.~R. Teixeira, B.~M. Praveen, G.~Nagaraju, Novel
  3d-flower shaped KTaO$_3$ perovskite for highly efficient photocatalytic and H$_2$
  generation ability, Scientific Reports 12 (2022) 10776.
\newblock \href {https://doi.org/10.1038/s41598-022-14590-3}
  {\path{doi:10.1038/s41598-022-14590-3}}.

\bibitem{Hussain2021}
M.~I. Hussain, R.~M. Khalil, F.~Hussain, A.~M. Rana, Dft-based insight into the
  magnetic and thermoelectric characteristics of XTaO$_3$ (X = Rb, Fr) ternary
  perovskite oxides for optoelectronic applications, International Journal of
  Energy Research 45 (2021) 2753--2765.
\newblock \href {https://doi.org/10.1002/er.5968} {\path{doi:10.1002/er.5968}}.

\bibitem{Gupta2022}
A.~Gupta, H.~Silotia, A.~Kumari, M.~Dumen, S.~Goyal, R.~Tomar, N.~Wadehra,
  P.~Ayyub, S.~Chakraverty, KTaO$_3$—the new kid on the spintronics block, Advanced Materials 34 (2022) 2106481.
\newblock \href {https://doi.org/10.1002/adma.202106481}
  {\path{doi:10.1002/adma.202106481}}.

\bibitem{Hussain2024}
M.~K. Hussain, R.~Paudel, B.~J. kahdum, S.~Syrotyuk, Optoelectronic and
  structural properties of bulk cubic perovskite RbTaO$_3$ and surface (001) for
  optoelectronic and spintronics applications, Materials Chemistry and Physics
  327 (11 2024).
\newblock \href {https://doi.org/10.1016/j.matchemphys.2024.129778}
  {\path{doi:10.1016/j.matchemphys.2024.129778}}.

\bibitem{Patra2023}
R.~Patra, P.~Dash, P.~K. Panda, P.~C. Yang, A breakthrough in photocatalytic
  wastewater treatment: The incredible potential of g-C$_3$N$_4$/titanate
  perovskite-based nanocomposites (8 2023).
\newblock \href {https://doi.org/10.3390/nano13152173}
  {\path{doi:10.3390/nano13152173}}.

\bibitem{Queisser1998}
H.~J. Queisser, E.~E. Haller,
  \href{https://www.science.org/doi/10.1126/science.281.5379.945}{Defects in
  semiconductors: Some fatal, some vital}, Science 281 (1998) 945--950.
\newblock \href {https://doi.org/10.1126/science.281.5379.945}
  {\path{doi:10.1126/science.281.5379.945}}.
\newline\urlprefix\url{https://www.science.org/doi/10.1126/science.281.5379.945}

\bibitem{Nakamura2003}
A.~Nakamura, K.~Matsunaga, J.~Tohma, T.~Yamamoto, Y.~Ikuhara, Conducting
  nanowires in insulating ceramics (2003).
\newblock \href {https://doi.org/10.1038/nmat920} {\path{doi:10.1038/nmat920}}.

\bibitem{Gao2018}
P.~Gao, S.~Yang, R.~Ishikawa, N.~Li, B.~Feng, A.~Kumamoto, N.~Shibata, P.~Yu,
  Y.~Ikuhara, Atomic-scale measurement of flexoelectric polarization at
  SrTiO$_3$ dislocations, Physical Review Letters 120 (6 2018).
\newblock \href {https://doi.org/10.1103/PhysRevLett.120.267601}
  {\path{doi:10.1103/PhysRevLett.120.267601}}.

\bibitem{Armstrong2021}
M.~D. Armstrong, K.~W. Lan, Y.~Guo, N.~H. Perry, Dislocation-mediated
  conductivity in oxides: Progress, challenges, and opportunities (6 2021).
\newblock \href {https://doi.org/10.1021/acsnano.1c01557}
  {\path{doi:10.1021/acsnano.1c01557}}.

\bibitem{Fang2024}
X.~Fang, Mechanical tailoring of dislocations in ceramics at room temperature:
  A perspective, Journal of the American Ceramic Society 107 (2024) 1425--1447.
\newblock \href {https://doi.org/10.1111/jace.19362}
  {\path{doi:10.1111/jace.19362}}.

\bibitem{Brunner2001a}
D.~Brunner, S.~Taeri-baghbadrani, W.~Sigle, M.~Ru, M.~Rühle, Surprising
  results of a study on the plasticity in strontium titanate, Journal of the
  American Ceramic Society 84 (2001) 1161--1163.
\newblock \href {https://doi.org/10.1111/j.1151-2916.2001.tb00805.x}
  {\path{doi:10.1111/j.1151-2916.2001.tb00805.x}}.

\bibitem{Sigle2006}
W.~Sigle, C.~Sarbu, D.~Brunner, M.~Rühle, Dislocations in plastically deformed
  SrTiO$_3$, Philosophical Magazine 86 (2006) 4809--4821.
\newblock \href {https://doi.org/10.1080/14786430600672695}
  {\path{doi:10.1080/14786430600672695}}.

\bibitem{Szot2006}
K.~Szot, W.~Speier, G.~Bihlmayer, R.~Waser, Switching the electrical resistance
  of individual dislocations in single-crystalline SrTiO$_3$, Nature Materials 5
  (2006) 312--320.
\newblock \href {https://doi.org/10.1038/nmat1614}
  {\path{doi:10.1038/nmat1614}}.

\bibitem{Rodenbucher2023}
C.~Rodenbücher, G.~Bihlmayer, C.~Korte, K.~Szot, Gliding of conducting
  dislocations in SrTiO$_3$ at room temperature: Why oxygen vacancies are strongly
  bound to the cores of dislocations, APL Materials 11 (2 2023).
\newblock \href {https://doi.org/10.1063/5.0126378}
  {\path{doi:10.1063/5.0126378}}.

\bibitem{Hirel2015}
P.~Hirel, A.~F. Mark, M.~Castillo-Rodriguez, W.~Sigle, M.~Mrovec, C.~Elsässer,
  Theoretical and experimental study of the core structure and mobility of
  dislocations and their relation to ferroelectric polarization in perovskite
  KnbO$_3$, Phys. Rev. B 92 (2015) 214101.
\newblock \href {https://doi.org/10.1103/PhysRevB.92.214101}
  {\path{doi:10.1103/PhysRevB.92.214101}}.

\bibitem{Khayr2024}
Issam Khayr, Sajna Hameed, Jakov Budić, Xing He, Richard Spieker, Ana Najev, Zinan Zhao, Li Yue, Matthew Krogstad, Structural properties of plastically deformed SrTiO$_3$ and KTaO$_3$, Phys. Rev. Materials 8 (2024) 124404.
\newblock \href {https://doi.org/10.1103/PhysRevMaterials.8.124404}
  {\path{doi:10.1103/PhysRevMaterials.8.124404}}.

\bibitem{Fang2024KTO}
X.~Fang, J.~Zhang, A.~Frisch, O.~Preuß, C.~Okafor, M.~Setvin, W.~Lu,
  \href{https://ceramics.onlinelibrary.wiley.com/doi/10.1111/jace.20040}{Room‐temperature
  bulk plasticity and tunable dislocation densities in KTaO$_3$}, Journal of
  the American Ceramic Society (7 2024).
\newblock \href {https://doi.org/10.1111/jace.20040}
  {\path{doi:10.1111/jace.20040}}.

\bibitem{Hirel2012}
P.~Hirel, M.~Mrovec, C.~Elsässer, Atomistic study of $\langle$110$\rangle$
  dislocations in strontium titanate, Acta Mater. 60 (2012) 329--338.
\newblock \href {https://doi.org/10.1016/j.actamat.2011.09.049}
  {\path{doi:10.1016/j.actamat.2011.09.049}}.

\bibitem{Marrocchelli2015}
D.~Marrocchelli, L.~Sun, B.~Yildiz, Dislocations in SrTiO$_3$: Easy to reduce
  but not so fast for oxygen transport, Journal of the American Chemical
  Society 137 (2015) 4735--4748.
\newblock \href {https://doi.org/10.1021/ja513176u}
  {\path{doi:10.1021/ja513176u}}.

\bibitem{Hirel2025}
P.~Hirel, P.~Cordier, P.~Carrez, $\langle 110\rangle \{\bar{1}10\}$ edge
  dislocations in strontium titanate: Charged vs neutral, glide vs climb, Acta
  Materialia 285 (2025) 120636.
\newblock \href {https://doi.org/10.1016/j.actamat.2024.120636}
  {\path{doi:10.1016/j.actamat.2024.120636}}.

\bibitem{Plimpton1995}
S.~J. Plimpton, \href{http://lammps.sandia.gov}{Fast parallel algorithms for
  short-range molecular dynamics}, J. Comp. Phys. 117 (1995) 1--19.
\newline\urlprefix\url{http://lammps.sandia.gov}

\bibitem{Pedone2006}
A.~Pedone, G.~Malavasi, M.~C. Menziani, A.~N. Cormack, U.~Segre, A new
  self-consistent empirical interatomic potential model for oxides, silicates,
  and silicas-based glasses, Journal of Physical Chemistry B 110 (2006)
  11780--11795.
\newblock \href {https://doi.org/10.1021/jp0611018}
  {\path{doi:10.1021/jp0611018}}.

\bibitem{Gale1997}
J.~D. Gale, Gulp: A computer program for the symmetry-adapted simulation of
  solids, J. Chem. Soc., Faraday Trans. 93 (1997) 629--637.
\newblock \href {https://doi.org/10.1039/A606455H}
  {\path{doi:10.1039/A606455H}}.

\bibitem{Xu2017}
Y.~Q. Xu, S.~Y. Wu, L.~J. Zhang, L.~N. Wu, C.~C. Ding, First-principles study
  of structural, electronic, elastic, and optical properties of cubic KNbO$_3$ and
  KTaO$_3$ crystals, Physica Status Solidi (B) Basic Research 254 (2017) 5.
\newblock \href {https://doi.org/10.1002/pssb.201600620}
  {\path{doi:10.1002/pssb.201600620}}.

\bibitem{Sepliarsky2005}
M.~Sepliarsky, A.~Asthagiri, S.~R. Phillpot, M.~G. Stachiotti, R.~L. Migoni,
  Atomic-level simulation of ferroelectricity in oxide materials, Current
  Opinion in Solid State and Materials Science 9 (2005) 107--113.
\newblock \href {https://doi.org/10.1016/j.cossms.2006.05.002}
  {\path{doi:10.1016/j.cossms.2006.05.002}}.

\bibitem{Hirth1982}
J.~P. Hirth, J.~Lothe, Theory of dislocations, 2nd edition, McGraw-Hill Publ.
  Co., New York, 1982.

\bibitem{Hirel2015Atomsk}
P.~Hirel,
  Atomsk: a tool for converting and manipulating atomic data files, Comput. Phys. Comm. 197 (2015) 212-219.
\newblock \href {https://doi.org/10.1016/j.cpc.2015.07.012}
  {\path{doi:10.1016/j.cpc.2015.07.012}}.

\bibitem{Momma2011}
K.~Momma, F.~Izumi,
  \href{http://scripts.iucr.org/cgi-bin/paper?S0021889811038970}{Vesta 3 for
  three-dimensional visualization of crystal, volumetric and morphology data},
  Journal of Applied Crystallography 44 (2011) 1272--1276.
\newblock \href {https://doi.org/10.1107/S0021889811038970}
  {\path{doi:10.1107/S0021889811038970}}.
\newline\urlprefix\url{http://scripts.iucr.org/cgi-bin/paper?S0021889811038970}

\bibitem{Stukowski2010}
A.~Stukowski, K.~Albe, Dislocation detection algorithm for atomistic
  simulations, Modelling Simul. Mater. Sci. Eng. 18 (2010) 25016.

\bibitem{Kelchner1998_PRB}
C.~L. Kelchner, S.~J. Plimpton, J.~C. Hamilton, No title, Phys. Rev. B 58
  (1998) 11085.

\bibitem{Jonsson1998}
H.~Jònsson, G.~Mills, K.~W. Jacobsen,
  Nudged elastic band method for finding minimum energy paths of transitions, in:
  Classical and Quantum Dynamics in Condensed Phase Simulations, World
  Scientific, 1998, pp. 385--404.
\newblock \href {https://doi.org/10.1142/9789812839664_0016}
  {\path{doi:10.1142/9789812839664_0016}}.

\bibitem{Henkelman2000}
G.~Henkelman, H.~Jónsson, Improved tangent estimate in the nudged elastic band
  method for finding minimum energy paths and saddle points, Journal of
  Chemical Physics 113 (2000) 9978--9985.
\newblock \href {https://doi.org/10.1063/1.1323224}
  {\path{doi:10.1063/1.1323224}}.

\bibitem{Weber2018}
M.~J. Weber, Handbook of Optical Materials, 2018.
\newblock \href {https://doi.org/10.1201/9781315219615}
  {\path{doi:10.1201/9781315219615}}.

\bibitem{Shigemi2006}
A.~Shigemi, T.~Koyama, T.~Wada, First-principles studies of various
  crystallographic phases and neutral atomic vacancies in KNbO$_3$ and KTaO$_3$, in:
  Physica Status Solidi (C) Current Topics in Solid State Physics, Vol.~3,
  2006, pp. 2862--2866.
\newblock \href {https://doi.org/10.1002/pssc.200669541}
  {\path{doi:10.1002/pssc.200669541}}.

\bibitem{Hu2022}
L.~Hu, D.~Sun, H.~Zhang, J.~Luo, C.~Quan, Z.~Han, K.~Dong, Y.~Chen, M.~Cheng,
  Growth, defects, mechanical, and optical properties of transparent KTaO$_3$
  single crystal, Journal of Materials Science: Materials in Electronics 33
  (2022) 13051--13063.
\newblock \href {https://doi.org/10.1007/s10854-022-08246-1}
  {\path{doi:10.1007/s10854-022-08246-1}}.

\bibitem{Benrekia2012}
A.~R. Benrekia, N.~Benkhettou, A.~Nassour, M.~Driz, M.~Sahnoun, S.~Lebègue,
  Structural, electronic and optical properties of cubic SrTiO$_3$ and KTaO$_3$: Ab
  initio and GW calculations, Physica B: Condensed Matter 407 (2012)
  2632--2636.
\newblock \href {https://doi.org/10.1016/j.physb.2012.04.013}
  {\path{doi:10.1016/j.physb.2012.04.013}}.

\bibitem{Bouafia2013}
H.~Bouafia, S.~Hiadsi, B.~Abidri, A.~Akriche, L.~Ghalouci, B.~Sahli,
  Structural, elastic, electronic and thermodynamic properties of KTaO$_3$ and
  NaTaO$_3$: Ab initio investigations, Computational Materials Science 75 (2013)
  1--8.
\newblock \href {https://doi.org/10.1016/j.commatsci.2013.03.030}
  {\path{doi:10.1016/j.commatsci.2013.03.030}}.

\bibitem{Kresse1996}
G. Kresse and J. Furthm\"uller, Efficient iterative schemes for ab initio total-energy calculations using a plane-wave basis set, Phys. Rev. B 54 (1996) 11169--11186.
\newblock \href{https://doi.org/10.1103/PhysRevB.54.11169}
  {\path{doi:10.1103/PhysRevB.54.11169}}.

\bibitem{KresseJoubert1999}
G. Kresse and D. Joubert, From ultrasoft pseudopotentials to the projector augmented-wave method, Phys. Rev. B 59 (1999) 1758--1775.
\newblock \href{https://doi.org/10.1103/PhysRevB.59.1758}
  {\path{doi:10.1103/PhysRevB.59.1758}}.

\bibitem{Blochl1994}
P.E. Bl\"ochl, Projector augmented-wave method, Phys. Rev. B 50 (1994) 17953--17979.
\newblock \href{https://doi.org/10.1103/PhysRevB.50.17953}
  {\path{doi:10.1103/PhysRevB.50.17953}}.

\bibitem{Perdew1996}
John P. Perdew, Kieron Burke, and Matthias Ernzerhof, Generalized Gradient Approximation Made Simple, Phys. Rev. Lett. 77 (1996) 3865--3868.
\newblock \href{https://doi.org/10.1103/PhysRevLett.77.3865}
{\path{doi:10.1103/PhysRevLett.77.3865}}.

\bibitem{Monkhorst1976}
Hendrik J. Monkhorst and James D. Pack, Special points for Brillouin-zone integrations, Phys. Rev. B 13 (1976) 5188--5192.
\newblock \href{https://doi.org/10.1103/PhysRevB.13.5188}
{\path{doi:10.1103/PhysRevB.13.5188}}.

\bibitem{Hirel2010}
P.~Hirel, P.~Marton, M.~Mrovec, C.~Elsässer, Theoretical investigation of
  \{110\} generalized stacking faults and their relation to dislocation
  behavior in perovskite oxides, Acta Materialia 58 (2010) 6072--6079.
\newblock \href {https://doi.org/10.1016/j.actamat.2010.07.025}
  {\path{doi:10.1016/j.actamat.2010.07.025}}.

\bibitem{Klomp2023}
A.~J. Klomp, L.~Porz, K.~Albe, The nature and motion of deformation-induced
  dislocations in SrTiO$_3$: Insights from atomistic simulations, Acta Materialia
  242 (2023).
\newblock \href {https://doi.org/10.1016/j.actamat.2022.118404}
  {\path{doi:10.1016/j.actamat.2022.118404}}.

\bibitem{Castillo-Rodriguez2010}
M.~Castillo-Rodríguez, W.~Sigle, Dislocation dissociation and stacking-fault
  energy calculation in strontium titanate, Scripta Materialia 62 (2010)
  270--273.
\newblock \href {https://doi.org/10.1016/j.scriptamat.2009.11.016}
  {\path{doi:10.1016/j.scriptamat.2009.11.016}}.

\bibitem{Xi2017}
J.~Xi, H.~Xu, Y.~Zhang, W.~J. Weber, Strain effects on oxygen vacancy
  energetics in KTaO$_3$, Physical Chemistry Chemical Physics 19 (2017)
  6264--6273.
\newblock \href {https://doi.org/10.1039/c6cp08315c}
  {\path{doi:10.1039/c6cp08315c}}.

\bibitem{Javaid2018}
F.~Javaid, K.~E. Johanns, E.~A. Patterson, K.~Durst, Temperature dependence of
  indentation size effect, dislocation pile-ups, and lattice friction in (001)
  strontium titanate, Journal of the American Ceramic Society 101 (2018)
  356--364.
\newblock \href {https://doi.org/10.1111/jace.15182}
  {\path{doi:10.1111/jace.15182}}.

\bibitem{Hirel2016b}
P.~Hirel, P.~Carrez, P.~Cordier, From glissile to sessile: Effect of
  temperature on (110) dislocations in perovskite materials, Scripta Materialia
  120 (2016) 67--70.
\newblock \href {https://doi.org/10.1016/j.scriptamat.2016.04.001}
  {\path{doi:10.1016/j.scriptamat.2016.04.001}}.

\end{thebibliography}
\bibliographystyle{elsarticle-num}

\end{document}